\documentclass[a4paper]{jpconf}
\usepackage{graphicx}
\usepackage{subcaption}
\usepackage{amsmath}
\usepackage{amssymb}
\usepackage{amsxtra}
\usepackage{cite}

\begin{document}
\title{Evolution of the cluster of primordial black holes within the Fokker-Planck approach}

\author{V D Stasenko$^1$, A A Kirillov$^1$}
    
\address{$^1$National Research Nuclear University MEPhI, (Moscow Engineering Physics Institute),
115409 Moscow, Russia}

\ead{stasenkovd@gmail.com}

\begin{abstract}
The calculation results of the evolution of the cluster of primordial black holes based on the Fokker-Planck equation with neglecting of the gas accretion onto black holes are presented. In addition, we consider how a massive black hole located within the cluster center affects on its evolution. Despite it creates an additional potential in the central region of the cluster and might capture surrounding black holes, a negligible growth rate of a central black hole was shown for 1 Gyr. Furthermore, we find a significant (approximately tenfold) expansion of the cluster.
\end{abstract}

\section{Introduction}
The idea that primordial black holes (PBHs) could form in the early Universe was proposed in \cite{1967SvA....10..602Z}. Then many theoretical mechanisms of the formation of PBHs were developed \cite{2010RAA....10..495K}. In our work, we consider those where PBHs can form a cluster. One of the possible mechanisms is collapse of closed domain walls \cite{2001JETP...92..921R, 2005APh....23..265K}. As a result of the collapse, the cluster gets the fractal structure with a large range of PBHs masses. In addition, a massive black hole appears in the cluster center. The mass of the central BH (CBH) is much less than the mass of the cluster. 
These clusters may resolve a lot of astrophysical and cosmological problems \cite{2014MPLA...2940005B, 2019EPJC...79..246B}, however, its observable features depend on PBHs distribution within the cluster at a given moment \cite{2019EPJC...79..246B, 2019JPhCS1390a2090K}.

In this work, we use the Fokker-Planck equation in order to study the dynamic of PBHs in the cluster. It is one of the classical approaches to investigate the evolution of star clusters (see \cite{2017ApJ...848...10V} and references within) which has much in common with the PBH clusters. However, unlike star clusters, clusters have a wide mass distribution and contain a CBH. These features lead to a slightly different rate of the evolution. Nevertheless, the evolution of the star cluster with a massive black hole was considered in the classical work \cite{1976ApJ...209..214B}. Afterward, the capture of stars by CBH was studied in terms of the loss cone in \cite{1977ApJ...211..244L, 1978ApJ...226.1087C}, and by now, this theory is being applied to galactic nuclei \cite{1991ApJ...370...60M, 2017ApJ...848...10V}.
In our work, we use the loss cone approach to describe the absorption of PBHs by CBH, but do not take into account the merges of other black holes due to the low mass of the cluster.


\section{The Fokker-Planck equation}

We use the approximation a relaxation time is much longer than an orbital period. We also assume that an isotropy exists in space of angular momentum. Thus, the distribution function $f(\textbf{\textit{r}}, \textbf{\textit{v}})$ of PBHs in the cluster depends on energy $E = v^2/2 + \phi(r)$ only, where $\phi(r)$ is gravitational potential. The Fokker-Planck equation in the energy space is follows \cite{2017ApJ...848...10V}:
\begin{equation} 
    \label{fp}
    \frac{\partial N_{i}}{\partial t} = \frac{\partial}{\partial E} \left (m_{i} D_{E}(E,f)\, f_{i} + D_{EE}(E,f)\, \frac{\partial f_i}{\partial E} \right) - \nu(E,f) N_{i},
\end{equation}
where $N_{i}$ is number density in the energy space $N_i(E) = 4 \pi^2 p(E) f_i(E)$, $f_{i}$ and $m_i$ is the distribution function and the mass of $i$-th type of the PBHs mass. $\nu N_i$ is the lose-cone term describing the capture of PBHs (with the angular momentum less than $L_{lc} = 2 c r_{g}$) by the CBH. The expressions for each term in \eqref{fp} can be found in \cite{2017ApJ...848...10V}. In order to solve equation \eqref{fp} as linear, the coefficients $D_{E}$, $D_{EE}$, $\nu$ are calculated from values of the distribution function from previous time steps.

All coefficients included in equation \eqref{fp} also depend on the gravitational potential which have the expression in the spherical symmetric case:
\begin{equation} 
    \label{poison}
    \phi(r) = - 4 \pi G \left ( \frac{1}{r} \int_0^r dr' \, r'^2 \rho(r') + \int_r^{\infty} dr' \, r' \rho(r') \right) - \frac{G M_{\bullet}}{r},
\end{equation}
where $M_{\bullet}$ is the mass of CBH and $\rho(r)$ is the density:
\begin{align}
    \rho(r) = 4 \pi \sum_{i} \int_{\phi(r)}^0 dE \, f_i(E) \sqrt{2(E - \phi(r))},
\end{align}
where the summation over all types of PBH masses is performed. In order to study the dynamical evolution of a self-gravitating system (e.g. PBHs cluster), it is necessary to solve both the Fokker-Planck equation \eqref{fp} and the Poisson equation \eqref{poison}. The technique of numerical integration of these equations was developed in \cite{1979ApJ...234.1036C, 1980ApJ...242..765C}.

\section{Results}
We use following dependence for the initial density profile for each type of PBHs:
\begin{equation}
    \rho_{i}(r) \propto \left ( \frac{r}{r_{0}} \right)^{-2} \left [ 1 + \left (\frac{r}{r_{0}} \right)^2 \right]^{-3/2},
\end{equation}
where $r_0 = 0.5$~pc. We choose the mass spectrum in the form \cite{2019EPJC...79..246B}:
\begin{equation}
    \frac{dN}{dM} \propto \frac{1}{M_{\odot}} \left ( \frac{M}{M_{\odot}} \right)^{-2},
\end{equation}
where the range of PBHs masses is from $10^{-2} M_{\odot}$ to $10 M_{\odot}$. We take the bin width such that the total masses of each component of PBHs are equal to each other. And we choose the mass of the CBH $M_{\bullet} = 100 M_{\odot}$ and the total cluster mass $M_\text{tot} = 10^5 M_{\odot}$.

\begin{figure}[!t]
	\centering
	\begin{subfigure}{0.48\textwidth}
		\includegraphics[width = \textwidth]{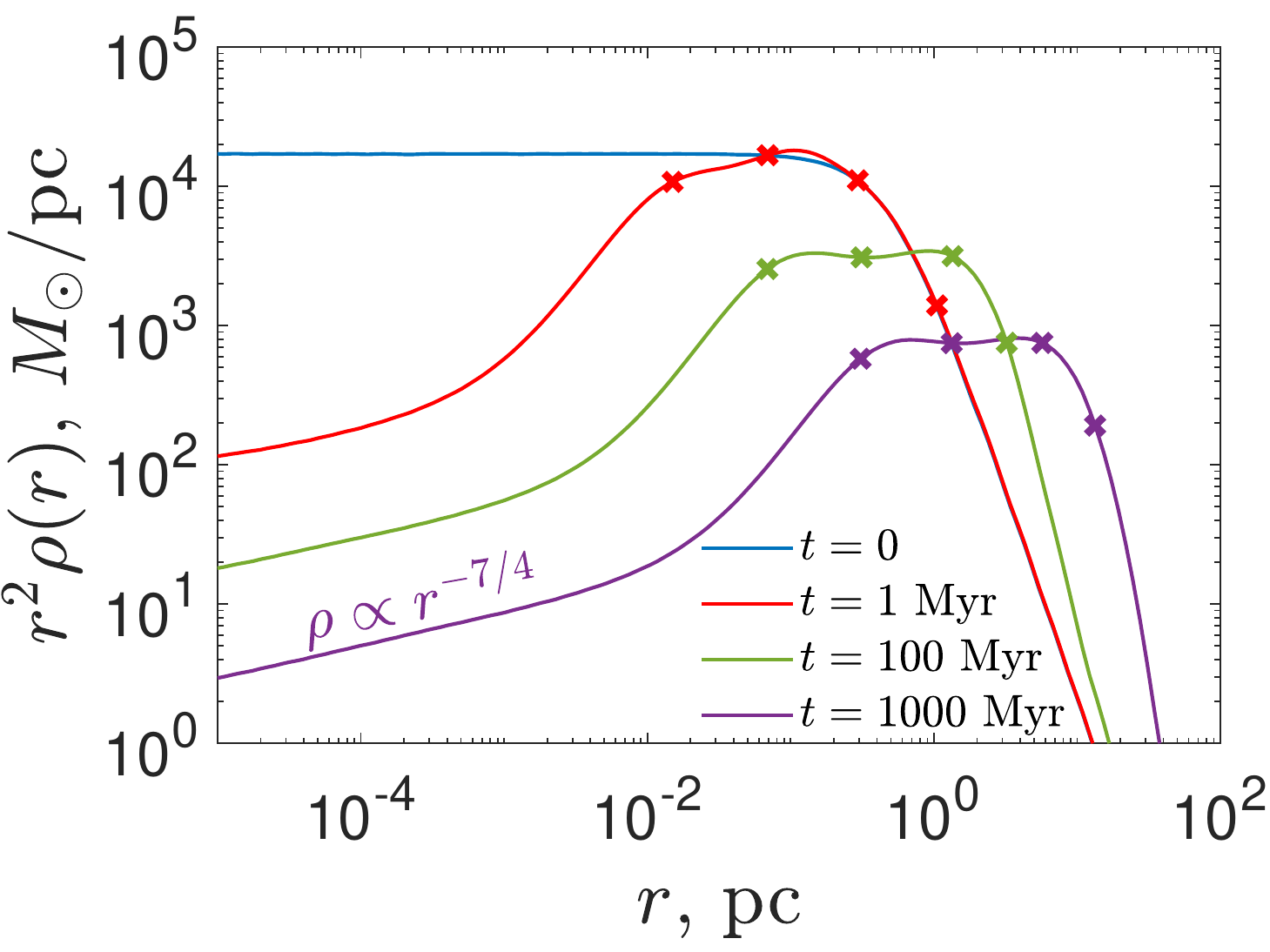}
		\caption{}
		\label{rho_tot}
	\end{subfigure}
	\hfil
	\begin{subfigure}{0.48\textwidth}
		\includegraphics[width = \textwidth]{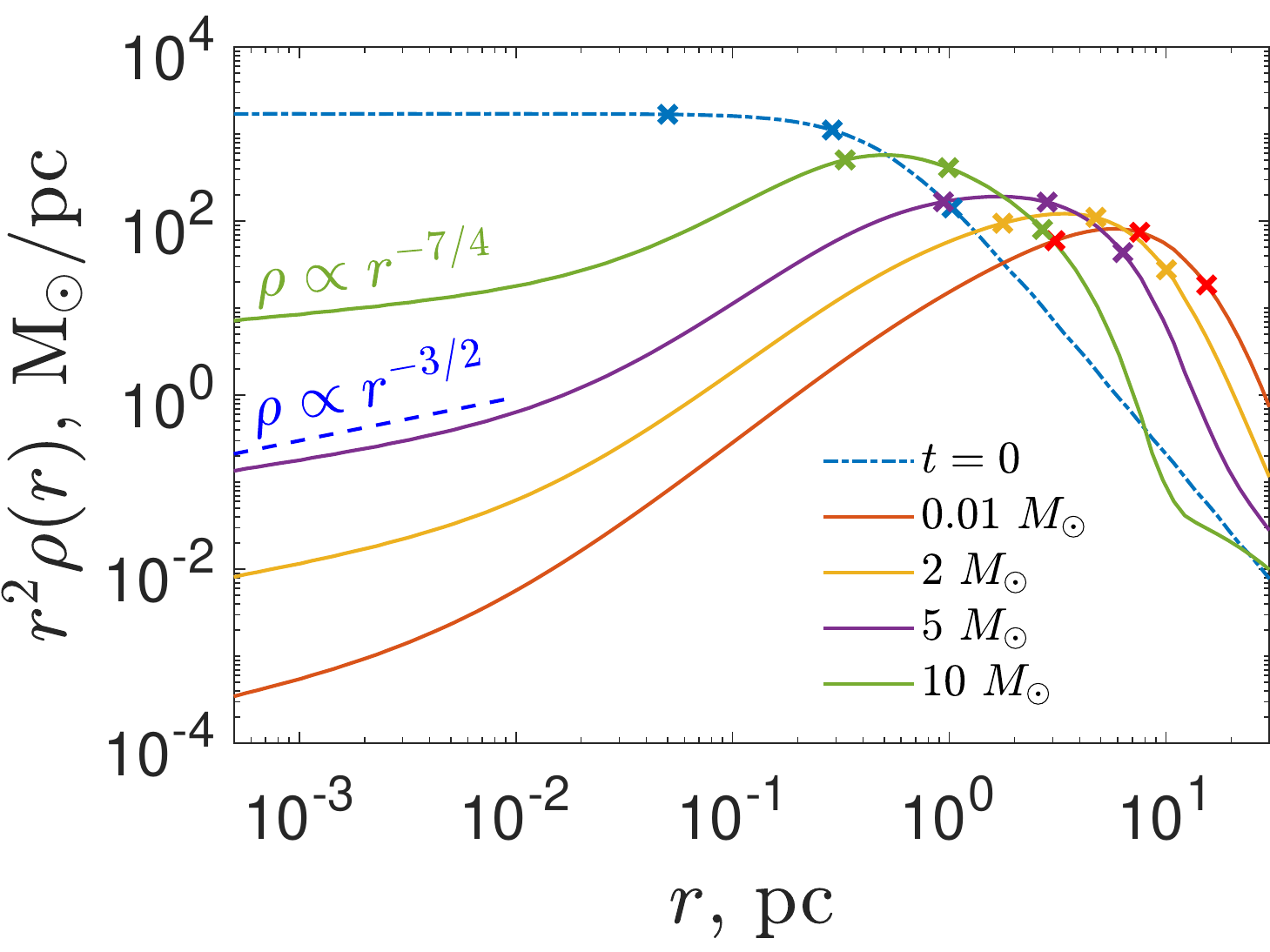}
		\caption{}
		\label{rho_comp_final}
	\end{subfigure}
	\caption{(a): the evolution of the density profile is shown. The crosses of each line from left to right correspond to radii containing 1, 10, 50 and 90\% of the total cluster mass, respectively. (b): the solid lines show the final density profiles of different PBHs mass types of the cluster at $t = 1$ Gyr. The dash-dotted line corresponds to the initial distribution (the same for all mass types). The crosses correspond to radii containing 10, 50 and 90\% of the total mass of each PBHs type.}
	\label{Rho_evolution}
\end{figure}

\begin{figure}[th!]
            \centering
            \includegraphics[width=0.5\textwidth]{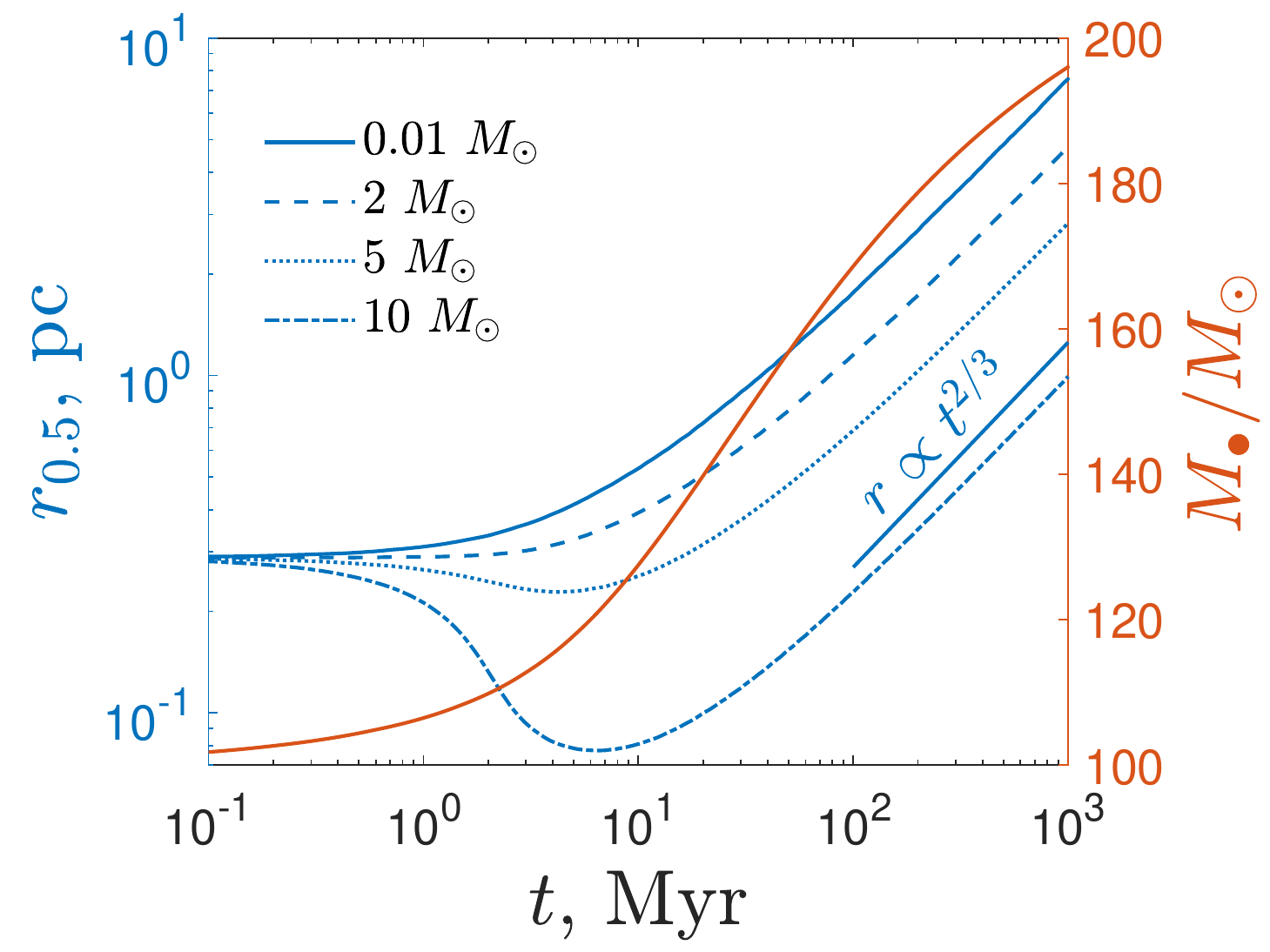}
            \caption{The blue lines (and the blue axis) show the evolution of the radius containing 50\% of the mass of each PBHs type. The red line (and the red axis) shows the growth of the central black hole.}
            \label{r50}
    \end{figure}
    
The evolution of the density profile is presented in figure \ref{rho_tot}. After small time ($\sim 1$~Myr) left, the cusp $\rho \propto r^{-7/4}$ is established in the central region of the cluster. Then, the cluster starts to expand and the most right red cross from the initial point $\sim 1$~pc goes to the purple cross $\sim 10$~pc at $t = 1$~Gyr. It is also seen, that $\sim 50 \%$ of total mass (first and third crosses) of the cluster approximately obeys the law for the density profile $\rho \propto r^{-2}$.

Figure \ref{rho_comp_final} illustrates the redistribution of the total mass of each PBHs type. It is seen at $t = 1$~Gyr, the radius containing 90\% of the heaviest PBHs mass corresponds the radius containing 10\% of the lightest PBHs mass. Thus, the significant changes have occurred in the structure of the cluster by the final moment of time, heavy PBHs are located closer to the cluster center than light ones. In fact, heavy PBHs are surrounded by light mass components of the cluster. It is also seen that density profile of heavy PBHs types have the behaviour in the central region $\rho \propto r^{-7/4}$. On the other hand, for small massive PBHs types the dependence has the form $\rho \propto r^{-3/2}$.

The evolution of the radius containing 50\% of the mass of each PBHs type and the growth of the CBH are presented in figure \ref{r50}. It is seen that at the first moment of the evolution, the heavy PBHs are compressed towards the center, but then the cluster expands according to $r \propto t^{2/3}$. CBH mass increases mainly within the first 100~Myr while the cluster has not expanded much yet.

\section*{Conclusion}

The dynamical evolution of the PBHs cluster is described within the framework of the orbit-averaged Fokker-Planck equation. We present the behavior of the density profile with time. It is obtained that the cluster size has increased by $\sim10$ times for 1~Gyr, and the CBH mass has increased by $\sim2$ times.

\section*{Acknowledgement}

The authors are grateful to K.~M.~Belotsky and S.~G.~Rubin for useful discussions. The work was supported by the Ministry of Science and Higher Education of the Russian Federation, project ``Fundamental properties of elementary particles and cosmology'' No 0723-2020-0041.

\section*{References}
\bibliographystyle{iopart-num}
\bibliography{iopart-num.bib}

\end{document}